\documentclass[12pt,a4]{article}

\usepackage{amssymb}
\usepackage{amsmath}
\usepackage{graphicx}

\begin{document}

\title{Relativistic persistent currents in ideal  Aharonov-Bohm rings}

\author{Ion I. Cot\u aescu, Doru-Marcel B\u alt\u a\c teanu and Ion  Cot\u{a}escu Jr. \\
{\small \it West University of Timi\c{s}oara,}\\
{\small \it V.  P\^{a}rvan Ave.  4, RO-300223 Timi\c{s}oara, Romania}}

\maketitle

\begin{abstract}
The exact solutions of the complete Dirac equation for fermions moving in  ideal Aharonov-Bohm rings  are used for deriving the exact expressions of the relativistic partial currents. It is show that as in the non-relativistic case  these currents can be related to the derivative of the fermion energy with respect to the flux parameter. A specific relativistic effect is the saturation of the partial currents for high values of the total angular momentum. Based on this property, the total relativistic persistent current at $T=0$ is evaluated giving its analytical expression and showing how this depend on the ring parameters.    
 
\end{abstract}

Keywords: Dirac equation; Aharonov-Bohm ring; persistent current.

\newpage

\section{Introduction}

The electronic effects in mesoscopic rings were studied by using the non-relativistic quantum mechanics \cite{B1}-\cite{B10}  based on the Schr\" odinger  equation with additional terms describing the spin-orbit interaction \cite{B11}-\cite{B15}. 

However, there are  nano-systems, as for example the graphenes,  where several relativistic effects can be observed in the electronic transport. These can be satisfactory explained  considering the electrons as  massless Dirac particles moving on honeycomb lattices \cite{N1}-\cite{Y2}.  Other Dirac materials are the topological insulators like $HgTe$ and $HgTe/CdTe$ quantum wells with low density and high mobility, in which the quantum spin Hall effect can be  realized \cite{Bu1,Bu2,Bu3}. 

Consequently,  many  studies  \cite{N1,N2}, \cite{B16}-\cite{B20} concentrate on the relativistic effects considering the electrons near the Fermi surface as being described by  the $(1+2)$-dimensional Dirac equation corresponding to a restricted three-dimensional Clifford algebra. However, in this manner one restricts simultaneously not only the orbital degrees of freedom but the spin ones too, reducing them to those of the $SO(1,2)$ symmetry.  

Under such circumstances, we believe that there are situations when it is convenient to use the {\em complete} $(1+3)$-dimensional Dirac equation  restricting the orbital motion, according to the concrete geometry of the studied system, but without affecting the natural spin degrees of freedom described by the $SL(2,{\Bbb C})$ group. Thus the polarization effects could be better pointed out. 

Nevertheless, the complete  Dirac equation was only occasionally used for investigating  some special problems of the fermions in external Aharonov-Bohm (AB) field as for example the spin effects in perturbation theory \cite{scat1,scat2,scat3}, the behaviour of the AB  fermions  in MIT cylinders \cite{MIT} and even  the AB dynamics using numerical methods.  

The persistent currents in AB quantum rings were recently studied starting with  a version of restricted Dirac equation involving a non-Hermitian term,  introduced by the orbital restrictions \cite{CP}. This  distort the results presented  therein  as well as other ones based on this approach,  even though these may outline new realistic effects as in Ref. \cite{Ghosh}. For this reason, we would like to continue this investigation here,  but considering  the {\em correctly restricted} Dirac equation, involving only Hermitian operators,  that can be obtained easily starting with a suitable restricted Lagrangian theory.

We discuss this topics showing first that the solutions of the  Dirac equation in the AB rings are determined as common eigenspinors of a  complete systems of commuting operators including the energy, total angular momentum and a specific operator analogous to the well-known Dirac spherical operator of the relativistic central problems \cite{TH}. These solutions can be normalized with respect to the relativistic scalar product obtaining thus the system of normalized fundamental solutions that allow us to write down the exact expressions of the relativistic partial currents and to derive the persistent ones.   

The relativistic partial currents we obtain here are related to the derivative of the relativistic energies as in the non-relativistic case but, in contrast, there appears a crucial difference: in the non-relativistic theory the partial currents are proportional to the angular momentum while in our approach the relativistic currents tend to {\em saturation} in the limit of high total angular momenta. For this reason we reconsider the problem of the relativistic persistent currents at $T=0$ proposing an approximative analytical formula that matches the numerical calculations with a satisfactory accuracy. 

The paper is organized as follows. In the second section we present the relativistic theory of the fermions in AB rings based on a suitable restriction of  the complete Dirac equation deducing the form of the normalized spinors. The next section is devoted to  the properties of the partial and persistent currents.  Finally we briefly present our conclusions.

\section{Dirac fermions in AB rings}

Let us consider a Dirac fermion of mass $M$  moving on a {\em ideal} ring of radius $R$  whose axis is oriented along the homogeneous and static external magnetic field $\vec{B}$ given by the electromagnetic potentials $A_0=0$ and $\vec{A}=\frac{1}{2}\vec{B}\land \vec{x}$ .  

The ideal ring is a one-dimensional manifold (without internal structure) embedded in the three-dimensional space according to the equations  $r=R$ and $z=0$,  written in cylindrical coordinates $(t,\vec{x}) \to (t, r, \phi, z)$ with the $z$ axis oriented along $\vec{B}$. Then, it is natural to assume that any field $\psi$ defined on this manifold depends only on the remaining coordinates $(t,\phi)$ such that $\partial_r\psi=0$ and $\partial_z\,\psi=0$. These restrictions give the kinetic term, 
\begin{equation}
{\cal S}_0=\int dt\, d\phi \,\left\{\frac{i}{2}\left[\overline{\psi}(\gamma^0\partial_t\psi +\gamma^{\phi}\partial_{\phi}\psi) -(\partial_t\overline{\psi}\gamma^0+\partial_{\phi}\overline{\psi}\gamma^{\phi})\psi \right]-M\overline{\psi}\psi\right\}\,,
\end{equation}
of the Dirac action ${\cal S}={\cal S}_0-\beta\int dt\, d\phi\, \overline{\psi}\gamma^{\phi}\psi$ in the mentioned external magnetic field, where $\overline{\psi}=\psi^{\dagger}\gamma^0$ and 
\begin{equation}\label{gamph}
\gamma^{\phi}=\frac{1}{R}(-\gamma^1\sin\phi+\gamma^2\cos\phi)
\end{equation}
is depending on $\phi$. The notation $\beta=\frac{1}{2}eBR^2$ stands for the usual dimensionless flux parameter (in natural units).

From this action we obtain  the correctly restricted Dirac equation, $E_D\psi=M\psi$,  with  the  new self-adjoint Dirac operator   
\begin{equation}\label{D2}
E_D=i\gamma^0\partial_t +\gamma^{\phi}(i\partial_{\phi}-\beta)+\frac{i}{2}\,\partial_{\phi}(\gamma^{\phi})\,,
\end{equation}
whose supplemental last term guarantees that $\overline{E}_D=E_D$. 
This operator commutes with the energy operator $H=i\partial_t$ and the third component, $J_3=L_3+S_3$, of the total angular momentum, formed by the orbital part $L_3=-i\partial_{\phi}$ the spin one $S_3=\frac{1}{2}\,{\rm diag} (\sigma_3,\sigma_3)$. Therefore, we have the opportunity to look for particular solutions of the form 
\begin{equation}\label{psi}
\psi_{E,\lambda}(t, \phi)=N\left(
\begin{array}{c}
f_1 e^{i\phi(\lambda-\frac{1}{2})}\\
f_2 e^{i\phi(\lambda+\frac{1}{2})}\\
g_1 e^{i\phi(\lambda-\frac{1}{2})}\\
g_2 e^{i\phi(\lambda+\frac{1}{2})}
\end{array}\right) e^{-iEt}\,,
\end{equation}
which satisfy the common eigenvalue problems, $E_D\psi_{E,\lambda}(t, \phi)=M\psi_{E,\lambda}(t, \phi)$ and
\begin{equation}
H\psi_{E,\lambda}(t, \phi)=E\psi_{E,\lambda}(t, \phi)\,, \quad  
J_3\psi_{E,\lambda}(t, \phi)=\lambda\psi_{E,\lambda}(t, \phi)\,,
\end{equation}
laying out the energy $E$ and  the angular quantum number $\lambda=\pm\frac{1}{2},\pm \frac{3}{2},...$ whose values are determined by the condition $\psi_{E,\lambda}(t, \phi+2\pi)=\psi_{E,\lambda}(t, \phi)$.

In this manner we separated the variables remaining with a system of algebraic equations  that  in the standard representation of the gamma matrices (with diagonal $\gamma^0$) reads
\begin{equation}
\left(
\begin{array}{cccc}
E-M &0&0 &\frac{i}{R}(\lambda+\beta)\\
0&E-M &-\frac{i}{R}(\lambda+\beta)&0\\
0&-\frac{i}{R}(\lambda+\beta)&-E-M&0\\
\frac{i}{R}(\lambda+\beta)&0&0&-E-M
\end{array}\right)\,\left(
\begin{array}{c}
f_1\\
f_2\\
g_1\\
g_2
\end{array}\right) =0\,.
\end{equation}
This system has non-trivial solutions only for the discrete values of energy  
\begin{equation}\label{ene}
E_{\lambda}=
\frac{1}{R}\left[M^2R^2+(\beta+\lambda)^2\right]^{\frac{1}{2}}\,,
\end{equation}
whose second terms encapsulate the AB effect. For each value $E_{\lambda}$ we find  two particular solutions  for which
\begin{equation}
\left(
\begin{array}{c}
f_1\\
f_2
\end{array}\right)= \xi_{\sigma} \,,
\end{equation} 
where $\xi_{\sigma}$ are the usual Pauli spinors of polarization $\sigma=\pm\frac{1}{2}$ with respect to the $z$ axis,
\begin{equation}
\xi_{\frac{1}{2}}=\left(
\begin{array}{c}
1\\
0
\end{array}\right)\,,\quad \xi_{-\frac{1}{2}}=\left(
\begin{array}{c}
0\\
1
\end{array}\right)\,.
\end{equation}
Thus,  we find that for $\sigma=\frac{1}{2}$ the spinors (\ref{psi})
take the form
\begin{equation}\label{psiUp0}
U_{\lambda}^+(t, \phi)=\frac{1}{2\sqrt{\pi E_{\lambda}R}}\left(
\begin{array}{c}
\sqrt{E_{\lambda}-M}\,    e^{i\phi(\lambda-\frac{1}{2})}\\
0\\
0\\
i\sqrt{E_{\lambda}+M}\,e^{i\phi(\lambda+\frac{1}{2})}
\end{array}\right) e^{-iE_{\lambda}t}\,,
\end{equation} 
while for $\sigma=-\frac{1}{2}$ we obtain the solutions
\begin{equation}\label{psiUm0}
U_{\lambda}^-(t,\phi)=\frac{1}{2\sqrt{\pi E_{\lambda}R}}\left(
\begin{array}{c}
0\\
\sqrt{E_{\lambda}-M}\,e^{i\phi(\lambda+\frac{1}{2})}\\
-i \sqrt{E_{\lambda}+M}\,e^{i\phi(\lambda-\frac{1}{2})}\\
0
\end{array}\right) e^{-iE_{\lambda}t}\,.
\end{equation}  
The normalization constants are fixed in accordance to the  relativistic scalar product
\begin{equation}
\langle \psi, \psi'\rangle=R\,\int_{0}^{2\pi}d\phi\,\psi^{\dagger}(t,\phi) \psi'(t,\phi)\,,
\end{equation} 
such that 
\begin{equation}
\langle U^{\pm}_{\lambda}, U^{\pm}_{\lambda'}\rangle=\delta_{\lambda,\lambda'}\,,\quad \langle U^{\pm}_{\lambda}, U^{\mp}_{\lambda'}\rangle=0\,.
\end{equation}

Hence we obtained a pair of fundamental solutions of the same energy and total angular momentum but which are not eigenspinors of the operators $L_3$ or $S_3$. Therefore, we may ask how these solutions can be defined as different eigenspinors of a new operator. The answer is obvious if we observe that the desired operator is  $K=2\gamma^0 S_3$ which satisfies
$K U^{\pm}_{\lambda}=\pm U^{\pm}_{\lambda}$. The conclusion is that the spinors  $U^{\pm}_{\lambda}$ are  common eigenspinors of the complete set of commuting operators $\{E_D,H,K,J_3\}$.
    
The operator $K$ introduced above is the analogous of the spherical Dirac operator $K_D=\gamma^0(2\vec{S}\cdot\vec{L}+1)$ that concentrates the angular variables of the Dirac equation in external fields with central symmetry \cite{TH}. Note that the genuine three-dimensional operator $K_D$ cannot be used here because of our dimensional reduction such that we must consider the simplified version $K$ \footnote{It is known that the forms of such operators depend on the number of space dimensions \cite{Dong}}. The eigenvalues of this operator give the polarization in the non-relativistic limit. For this reason we keep this terminology considering that the eigenvalues $\kappa=\pm1$ of the operator $K$ define the fermion polarization with respect to the direction of the magnetic field $\vec{B}$.

\section{Relativistic currents}

Using the above results we can calculate the exact relativistic expressions of the partial  currents on quantum rings, pointing out the difference between the genuine relativistic theory and  the non-relativistic one. We show that in the relativistic approach the partial current tends to saturation for increasing $\lambda$ such that the persistent currents at $T=0$ will get new properties.  

\subsection{Partial currents}

Let us start with the quantum rings where the states of the fermions of energy $E_\lambda$ are  described by the normalized linear combinations
\begin{equation}
\psi_{\lambda}=c_+U^+_{\lambda}+c_-U^-_{\lambda}\,, \quad |c_+|^2+|c_-|^2=1\,,
\end{equation} 
for which  the expectation value of the polarization operator reads,
\begin{equation}
\langle \psi_{\lambda}, K\psi_{\lambda}\rangle =|c_+|^2-|c_-|^2\,.
\end{equation}

The partial  currents (of given $\lambda$) coincide in this case with their  densities,     
$I_{\lambda}=R\,\overline{\psi}_{\lambda}\gamma^{\phi}\psi_{\lambda}$,
that can be calculated with the help of the matrix (\ref{gamph}). Then, observing that 
\begin{equation}\label{inter}
{\overline{U}_{\lambda}^{\pm}}(t,\phi)\gamma^{\phi}U_{\lambda}^{\mp}(t,\phi)=0\,,
\end{equation}
we obtain the {\em partial}  current of a fermion of energy $E_{\lambda}$ as
\begin{equation}
I_{\lambda}=|c_+|^2 I^+_{\lambda}+|c_-|^2I_{\lambda}^-=\frac{1}{2\pi R^2}\frac{\beta+\lambda}{E_{\lambda}}=\frac{1}{2\pi}\frac{\partial E_{\lambda}}{\partial\beta}\,,
\end{equation}
since  $I^{\pm}_{\lambda}=R\,{\overline{U}_{\lambda}^{\pm}}(t,\phi)\gamma^{\phi}U_{\lambda}^{\pm}(t,\phi)=I_{\lambda}$. Thus we find that the partial currents are independent on polarization being related to energies in a similar manner as in the non-relativistic theory.

The  exact relativistic expressions of the partial  currents  we obtained here depend only on two dimensionless parameters $\nu=\beta+\lambda$ and  $\mu=MR$ (or $MRc/\hbar$ in usual units) that are the arguments of the auxiliary function $\chi$ defined as
\begin{equation}\label{Irel}
I_{\lambda}=\frac{1}{2\pi R}\, \chi (\mu,\nu)\,, \quad \chi(\mu,\nu)=\frac{\nu}{\sqrt{\mu^2+\nu^2}}\,.
\end{equation} 
This function  has the remarkable  asymptotic behaviour 
\begin{equation}
\lim_{\nu\to\pm\infty}\chi(\mu,\nu)=\pm 1\,,
\end{equation}
which shows that the relativistic partial  currents tend to {\em saturation} for large values of $\lambda$.  Moreover, for small values of $\nu$ we can expand
\begin{equation}\label{cucu}
\chi(\mu,\nu)= \frac{\nu}{\mu}+O(\nu^3)\,.
\end{equation}
Note that the non-relativistic limit recovers the well-known behaviours
\begin{equation}
E_{\lambda}-M \to \tilde E_{\lambda}=\frac{\nu^2}{2 R\mu}\,, \quad I_{\lambda}\to \tilde I_{\lambda}=\frac{1}{2\pi R}\frac{\nu}{\mu}=\frac{1}{2\pi}\frac{\partial \tilde E_{\lambda}}{\partial \nu}\,.
\end{equation} 

Hereby we conclude  that the principal difference is that the relativistic partial currents (\ref{Irel}) are saturated  while in the non-relativistic case we do not meet this effect since the function $\chi(\mu,\nu)$ is replaced then by the  linear function $\frac{\nu}{\mu}$  that  is just its  tangent in $\nu=0$  as we deduce from Eq. (\ref{cucu}). This result was previously outlined in Ref. \cite{Ghosh} but based on the non-Hermitian Dirac equation of Ref. \cite{CP}.
Obviously, the correct saturation effect is given by the expression of the partial currents  (\ref{Irel}) derived here.  
 
Note that the non-relativistic approximation can be used with a satisfactory accuracy  only in the domain where the function $\chi(\mu,\nu)$ is  approaching to the linear function $\frac{\nu}{\mu}$. Our numerical evaluations show that  in the domain  $-\frac{1}{2}\mu<\nu<\frac{1}{2}\mu$  the difference $|\chi(\mu,\nu)-\frac{\nu}{\mu}|$ is satisfactory small remaining less than  0.05.   In addition,  we estimate that for $|\nu|\ge 5\mu$ the current is approaching to its saturation value since $ |\chi(\mu,\pm 5\mu)|= 0.98058$.

\subsection{Relativistic persistent currents}

The above results allow us to derive the total persistent current at $T=0$ in a semiconductor ring of parameter $\mu$ having a even number of electrons $N_e$ fixed by the Fermi-Dirac statistics. For the mesoscopic rings with $R=100 {\rm nm}$ the parameter $\mu$ is of the order $10^3 - 10^5$. For example, in a  $InSb$ ring of this radius,  the effective electron mass is $M=m^*_e=0.0135\, m_e$  \cite{Vag} such that  $\mu=3495$. This  seems to be the minimal value of $\mu$  obtained so far but  it is possible to obtain smaller values  in further experiments with mesoscopic rings with $R< 100 {\rm nm}$ or even with nano-rings having $R\sim 10 {\rm nm}$.   According to our estimation, the relativistic effects may be measurable for $\mu<10^3$ which means that the actual experiments  are approaching to this threshold which could be reached soon.

In all these cases the flux parameter $\beta$ remains very small (less than $10^{-8}$) such that we can neglect the terms of the order $O(\beta^2)$ of the Taylor expansions of our functions that depend on  $\nu=\lambda+\beta$.  The total persistent current at $T=0$  is given by the sum  
\begin{equation}
I= \sum_{\lambda=-\lambda_F}^{\lambda_F}I_{\lambda}= \sum_{\lambda=\frac{1}{2}}^{\lambda_F}\left(I_{\lambda}+I_{-\lambda}\right)=\frac{1}{2\pi R}\sum_{\lambda=\frac{1}{2}}^{\lambda_F}[\chi(\mu,\lambda+\beta)+\chi(\mu,-\lambda+\beta)]
\end{equation}
over all the allowed polarizations, $\lambda=\pm\frac{1}{2},\pm\frac{3}{2},...,\pm\lambda_F$ where $\lambda_F=\frac{1}{2}(N_e-1)$.  Furthermore,  by using the expansion
\begin{equation}
\chi(\mu,\lambda+\beta)+\chi(\mu,-\lambda+\beta)=2 j(\mu,\lambda) \beta +O(\beta^3)\,,
\end{equation}
where
\begin{equation}
j(\mu,\lambda)=\frac{\mu^2}{(\mu^2+\lambda^2)^{\frac{3}{2}}}\,,
\end{equation}
we arrive at the relativistic persistent currents,  
\begin{equation}\label{Ic}
I=c(\mu) I_{max}\,,\quad I_{max}=\frac{\beta}{\pi R}\,, \quad c(\mu)=\sum_{\lambda=\frac{1}{2}}^{\lambda_F}j(\mu,\lambda)\,,
\end{equation}
that can be calculated numerically on computer for any concrete value of $\mu$.

{ \begin{figure}
    \centering
    \includegraphics[scale=0.37]{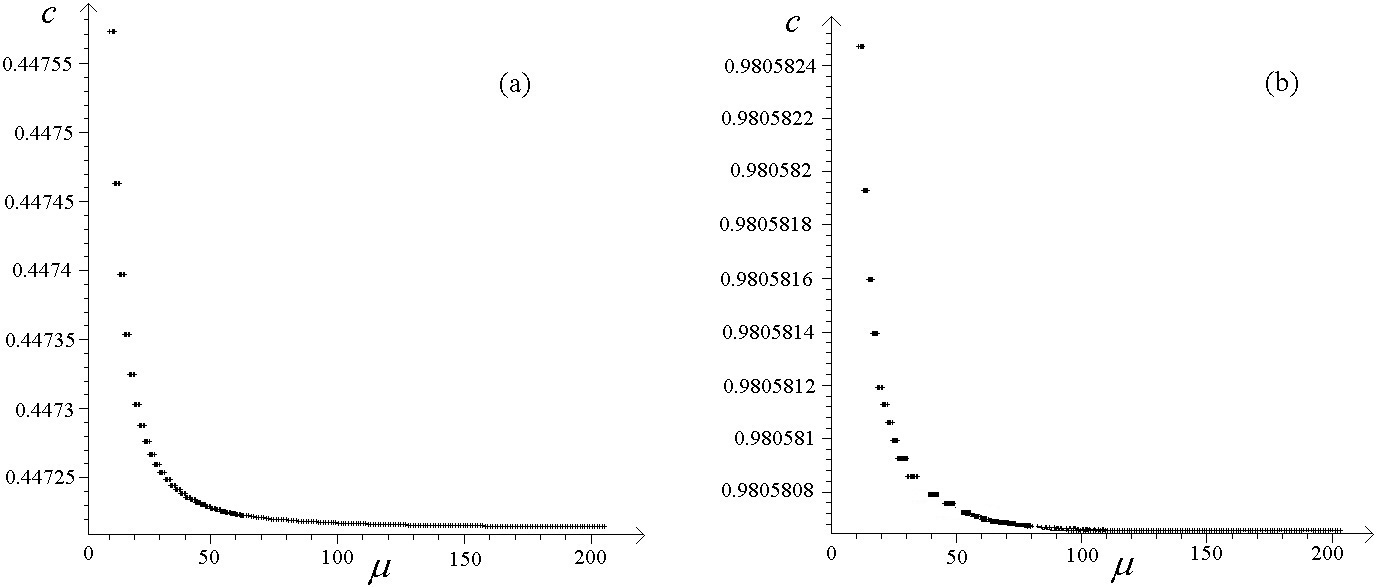}
    \caption{The function $c(\mu)$ versus $\mu$ calculated for $\lambda_F=0.5\mu$ (a) and $\lambda_F=5 \mu$  when the co-domain is very narrow, $\sim 10^{-5}$  (b).  }
  \end{figure}}

The function $j(\mu,\lambda)$ is simple reaching its maximal value 0.7698 for $\mu=\frac{1}{\sqrt{2}}$ and $\lambda=\frac{1}{2}$ and  decreasing  then monotonously  to zero when $\mu$ and $\lambda$ are increasing to infinity.  This behaviour is a direct consequence of the saturation of the partial currents that compensate each other in the saturation zone where $I_{\lambda}+I_{-\lambda} \to 0$.   These simple monotony and smoothness properties of the function $j(\mu,\lambda)$ lead to nice results concerning the values of the sum (\ref{Ic}c) when we compute all the allowed contributions. Our numerical examples show that when $\mu$ is increasing then the functions $c(\mu)$ are monotonously decreasing tending to an asymptotic value (as in Fig. 1). Consequently, in the asymptotic zone, $\mu>100$, we can use the following approximation
\begin{equation}\label{Intc}
c(\mu)\simeq \int_{0}^{\lambda_F}j(\mu,\lambda)d\lambda= \frac{\lambda_F}{\sqrt{\mu^2+\lambda_F^2}}\,,
\end{equation}
giving the definitive formula of the relativistic persistent currents
\begin{equation}\label{final}
I=\frac{k}{\sqrt{1+k^2}}\, I_{max}\,, \quad k=\frac{\lambda_F}{\mu}\simeq \frac{N_e}{2\mu}\,,
\end{equation} 
that reproduces the numerical results with a satisfactory accuracy (under $10^{-5}$). 
Note that the non-relativistic persistent current, that in our notation reads $\tilde I=k I_{max}$, represents a good approximation of Eq. (\ref{final}) only for small values of $k$ (say $k<0.2$) for which we can use the approximation $k(1+k^2)^{-\frac{1}{2}}=k+ O(k^3)\simeq k$. 

\section{Concluding remarks}

We outlined here the relativistic theory of the Dirac fermions in ideal AB rings. We found the complete system of the commuting operators determining the fundamental solutions that  contains the new operator $K$ which is the analogous of the Dirac spherical operator of the central problems. Thus we derived  the polarized spinors that can be normalized with respect to the relativistic scalar product. The corresponding relativistic partial currents have an interesting behaviour for  large values of  polarization,  tending to the saturation value $(2\pi R)^{-1}$ which depends only on the ring radius.  This property allowed us to derive the  relativistic persistent current at $T=0$ giving the closed expression  (\ref{final}). 

In our opinion,  the form of the relativistic partial current is suitable for estimating the total currents even for $T>0$ by using the Fermi-Dirac statistics. This is because in the saturation domain  the currents have very small contributions such that the sums become satisfactory convergent and can be easily performed on computers or estimated by appropriate integrals as in Eq. (\ref{Intc}). 

Finally, we observe that the relativistic theory based on the complete Dirac equation is able to offer new interesting results in investigating systems of low energy as those of the solid state physics that seemed to be destined exclusively to the non-relativistic quantum mechanics.   

\subsection*{Acknowledgments}

I. I. Cot\u aescu  is supported by a grant of the Romanian National Authority for Scientific Research, Programme for research-Space Technology and Advanced Research-STAR, project nr. 72/29.11.2013 between Romanian Space Agency and West University of Timisoara.

D.-M. B\u alt\u a\c teanu is supported by the strategic grant POSDRU /159/1.5/S /137750, Project “Doctoral and Postdoctoral programs support for increased competitiveness in Exact Sciences research”, cofinanced by the European Social Fund within the Sectoral Operational Programme Human Resources Development 2007 - 2013.

\end{document}